\documentclass[aps,showpacs,preprintnumbers,amsmath,amssymb,twocolumn, tightenlines]{revtex4}

\usepackage{graphicx}
\usepackage{bm}
\usepackage{amsmath}
\usepackage{dcolumn}
\usepackage{bm}

\newcommand{\be}{\begin{eqnarray}}
\newcommand{\ee}{\end{eqnarray}}

\begin{document}
\title{Bose-Einstein Condensation of strongly interacting bosons:\\
from liquid ${}^4$He to QCD monopoles}

\author{ Marco Cristoforetti$^1$ and Edward V. Shuryak$^2$ }

\affiliation{ $^1$Physik Department, Technische Universit\"at M\"unchen, D-85747 Garching, Germany\\
$^2$Department of Physics and Astronomy, State University of New York, 
Stony Brook NY 11794-3800, USA
}

\date{\today}

\vspace{0.1in}
\begin{abstract}
Starting from classic work of Feynman on the $\lambda$-point of liquid Helium, we show that his idea of universal action per particle at the BEC transition point
is much more robust that it was known before. Using a simple ``moving string model'' for supercurrent and calculating the action, both semiclassically and numerically,
we show that  the critical action is the same for noninteracting and strongly interacting systems such as liquid  ${}^4$He. Inversely, one can obtain accurate dependence of critical temperature on density: one important consequence is that  high density (solid) He cannot be  a BEC state of He atoms, with upper density accurately matching the observations. 
We then use this model for the deconfinement phase transition of QCD-like gauge theories, treated as BEC of (color)magnetic monopoles. We start with Feynman-like approach without interaction, estimating the monopole mass at $T_c$. Then we include monopole's Coulomb repulsion, and formulate a relation between the mass, density and coupling which should be fulfilled at the deconfinement point. We end up proposing various ways to test on the lattice whether it is indeed the BEC point for monopoles. 
  \end{abstract}

\maketitle

\section{Introduction}

The goals of this paper are two-fold. The {\em first goal} is rather general: to get better qualitative understanding of the parameters controlling the transition between the ``normal" matter and its Bose condensed versions, for strongly interacting bosons which may be in form of a liquid (fluid-superfluid transition well known for ${}^4$He) or solid (solid-supersolid transition yet to be found). Although there are high quality Monte Carlo numerical results for ${}^4$He and many other systems, we think  the 
 universal  condensation criterion is still very much needed, as performing numerical simulations is not trivial in each new settings.  
 
As it will be explained in detail below, we will follow 50-year-old   Feynman theory of Bose condensation \cite{fey1,fey2}, in which he introduced the notion of the critical value for the jump amplitude $y_c$ or the critical action $y_c=exp(-S_c)$.  When Feynman realized that his simple treatment (evaluation of only the kinetic energy part of the action) needs correction, he simply introduced an ``effective mass" thinking that some extra matter is incorporated into exchange motion. We think instead, that other particles (except  the ones in the exchanged polygon) have very little chance to move. Instead, the jump amplitude should  be correctly evaluated, with the interaction term included. We thus revive Feynman's idea, using a simple model of particle motion --the ``moving string model'' -- which can be studied either semiclassically or numerically of the smallest-action paths which particles should follow during their exchanges.

Our {\em second goal} is very far from atomic systems: it is related with the {\em deconfinement phase transition in QCD} and related gauge theories.
``Dual superconductivity" or BEC of certain magnetic objects in the vacuum of these theories
 were proposed to be responsible for confinement
 by t'Hooft  and Mandelstam \cite{t'Hooft-Mandelstamm}.  As shown in refs  \cite{Baker:1991bc,Ripka:2003vv},
 confining strings (electric flux tubes)   can be well described by effective models, making them dual versions
   of the Abrikosov flux tubes in  superconductors \cite{Bali}. 
   
   As Feynman did in 1950's, we would however approach the problem from the ``normal" phase 
 above the deconfinement phase transition  called  ``Quark Gluon Plasma" (QGP).
At very high $T$ one can view QGP as a  plasma made of ``electric'' quasiparticles, quarks and gluons. However
at lower $T$   it becomes a ``dual plasma" containing not only electrically charged quasiparticles but also magnetically charged objects --
monopoles and dyons.  As it was pointed out in \cite{Liao:2006ry}, with a decrease of the temperature and increase of the electric coupling constant  one expects a gradual shift from electric to magnetic dominance, with large density of (color-magnetic) monopoles near  $T_c$.  
Recent lattice studies, in particularly \cite{D'Alessandro:2007su}, have discovered many
important details about such monopoles. They indeed found rather high densities of them close to $T_c$,
as well as clustering behavior similar to BEC. They also measured equal-time density correlators of like and unlike monopoles (Fig.~5 of  \cite{D'Alessandro:2007su}) which clearly shows the peaks characteristic for strongly coupled ionic liquids. The estimated magnetic Coulomb coupling does  indeed show running opposite
to electric coupling predicted in \cite{Liao:2006ry}: see review \cite{Shuryak:2008eq}.
So, as QCD deconfinement transition is  interpreted as BEC of monopoles,  in the second part of the paper we will apply Feynman's universal action idea
to constrain the properties of these monopoles (mass, density and coupling constant) for BEC to become possible at the deconfinement point.


\begin{figure}[!ht]
\vspace{0.75cm}
\includegraphics[width=4cm]{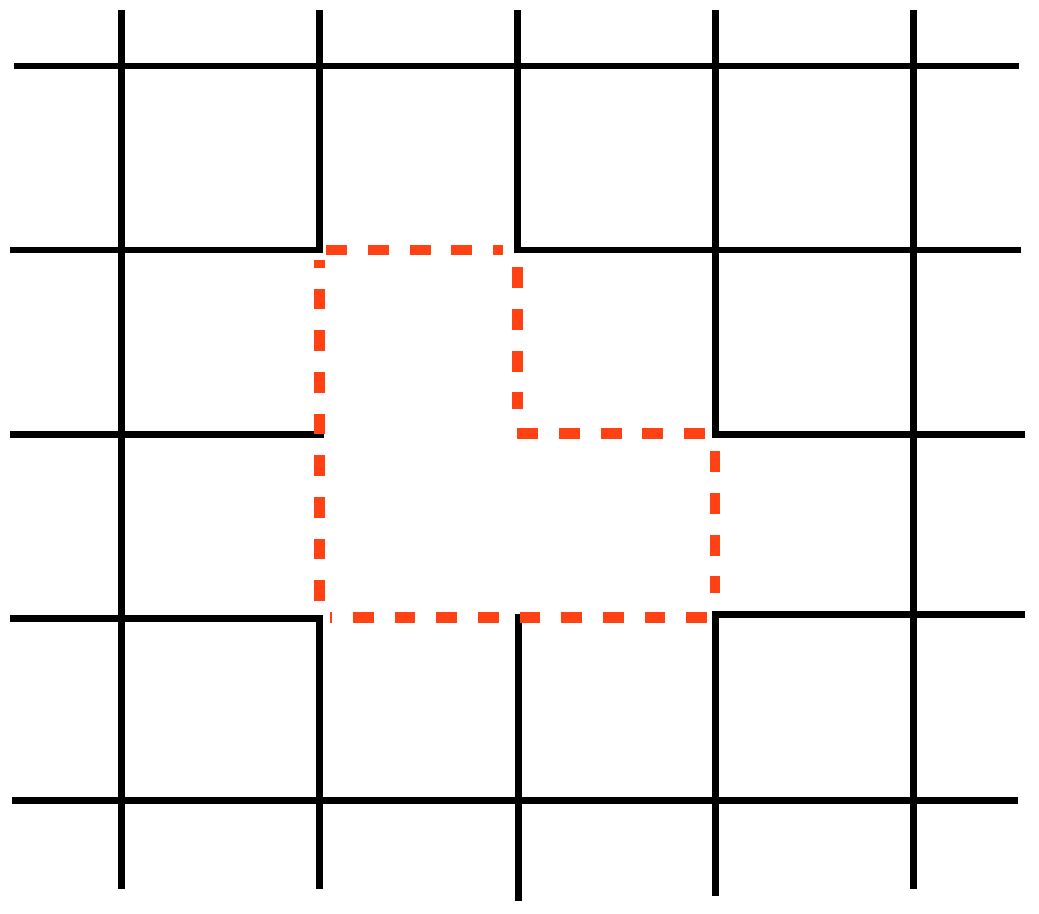}
\caption[h]{\label{fig_lattf} Example of polygon due to particle exchange}
\end{figure}

\section{Reviving the Feynman's theory of the Helium $\lambda$-point, 55 years later}\label{sec:feynm}
\subsection{Feynman's theory}
Feynman's  starting point is the thermal partition function in the form
\be
	Z&=&e^{-\frac{F}{T}}\\\nonumber
	&=&\frac{1}{N}\sum_P\int\Big(\frac{m^*T}{2\pi\hbar}\Big)^{3N/2}\exp\Big[-\frac{m^*T}{\hbar^2}\sum_i(\underline{R}_i-P\underline{R}_i)^2\Big]\\\nonumber
	&&\rho(\underline{R}_1...\underline{R}_N)\textrm{d}^3\underline{R}_1...\textrm{d}^3\underline{R}_N
\ee
where $m^*$ is the effective mass of a He atom, the sum is done over permutation $P$ of the particle coordinates, the function $\rho$ includes effects of the interparticle interactions and the exponent contains additional action due to kinetic energy. His main idea is that the function $\rho$ can be inferred from general properties of the liquid, its quasi-ordered local structure with peaked distribution over interparticle distances at some nearest-neighbor value $d$ (in the case of cubic lattice $d$ is the lattice spacing). Then the relative magnitude of a term with permutation of $n$ atoms would be proportional to the $n$-th power of the "jump amplitude" which is approximated as
\be\label{eq:yf}
	y_F=\exp\Big[-\frac{m^*Td^2}{2\hbar^2}\Big]
\ee
with some combinatorial prefactors, describing a number of corresponding to non-crossing polygons in the $3$d lattice. The divergence of this sum, at the parameter approaching some critical value $y_F\rightarrow y_c$ should indicate the presence of "infinite cluster", the signature of Bose condensation. The condition for this, Feynman argues, is $y_c\approx1/s$ where $s$ is the growth factor in the number of polygons when $n$ is increased by one unit. Feynman mentioned that he expected $y_c=1/4$ -- $1/3$.

The combinatoric problem to find the correct critical value of the Feynman parameter, was studied in details by Kikuchi et al.~\cite{kik}. They found a critical action $S_c\approx 1.9$. We can consider the distance $d$ of a "jump" fixed to the position of the nearest neighbor maximum in the static correlation function $g(r)$ for liquid He, which is $d\approx3.5$ \AA. If this value is used and the critical action is taken to be the one obtained by Kikuchi, in order to recover the correct position of the $\lambda$-point $T_c=2.17$, the Helium mass is changed to an effective value $m^*=1.64m_{\textrm{He}}$. Otherwise, using the physical Helium mass we obtain $T_c=3.57$ which is, of course $64\%$ bigger than the real critical temperature. In the eighties Elser~\cite{elser} determined numerically that the critical action should be smaller than the values obtained in~\cite{kik}, $S_c\approx 1.44$. Using this new parameter the predicted critical temperature reduces to $T_c=2.72$, closest to the physical temperature.

In the Feynman picture all the effects of the potential are absorbed into an effective mass $m^*$ that appears in the kinetic part of the partition function. In this letter we try to study the He $\lambda$-point, starting from the idea of Feynman, but including the effects due to the potential at a mean field level. In our calculations we have considered the Aziz potential HFDHE2 \cite{aziz}. We will show which in this context an estimate for the Helium critical temperature can be obtained within an error of $5\%$ if we assume that the critical action will be the same than the free case, that, as we will show, can be estimated to $S_c=1.655$ for a cubic lattice.

\begin{figure}[!ht]
\vspace{0.75cm}
	\includegraphics[width=5cm]{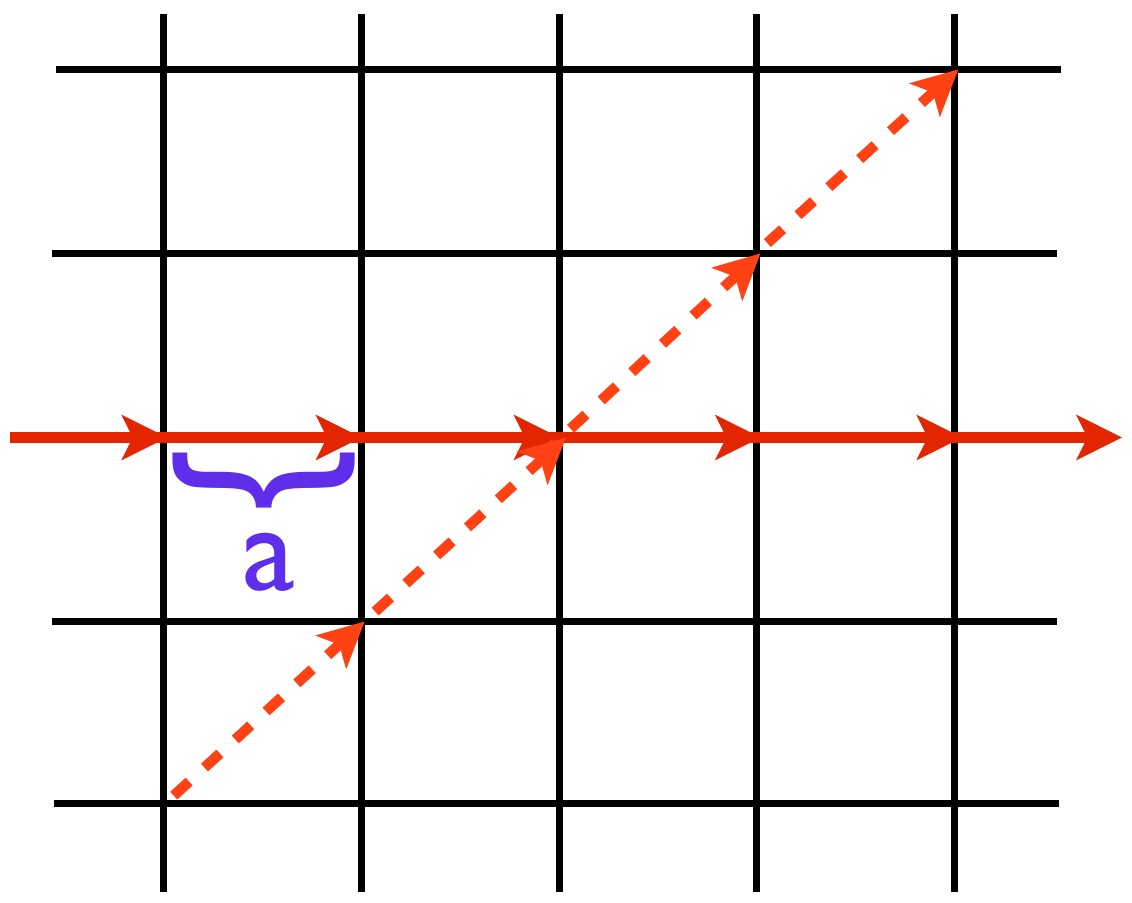}
\caption[h]{\label{fig_latt} $z=0$ of the three dimensional lattice: the red continuous arrows identify the moving atoms. The dashed arrows represents the diagonal trajectory of monopoles discussed in Sec.~\ref{sec:mon}}
\end{figure}

Following Feynman we put the Helium atoms filling a cubic lattice and consider two different states for the system: the first where the atoms are static on the lattice sites,  the second where the atoms on a line of the lattice are moving coherently in a given direction, for example along the $x$, see  Fig.~\ref{fig_latt}. If we consider an infinite system we can consider that this line is a side of a polygon with infinite number of atoms. Another difference with the Feynman's work is that in our case the atoms lying on the moving line are not restricted to move on the lattice link: what they must do is to jump from one site to the following on the same line, but in between they move following the three-dimensional equation of motion imposed by the geometry of the selected configuration.

The appearance of such a kind of particle cluster, as proposed by Feynman, is connected to the transition into the superfluid phase. Confirmation of this hypothesis come also from Path Integral Monte Carlo (PIMC) calculations, which, at today, represent the only way to determine without any approximation, the Helium critical temperature. In this framework, the transition point is fixed exactly from the appearance of atoms involved in an exchange which winds around the simulation box. 

The idea of Feynman was to consider the ratio between the partition function of the system where permutations are possible with respect to the "Boltzmann" case where particles do not permute. In our case we are interested to understand the properties of the particular system where there is exactly one line involved in permutation with respect to the Boltzmann system. 

Systems at finite temperature can be studied using path integral techniques in the Euclidean space, where $t\rightarrow -i\tau$. 

In general the partition function for bosonic interacting particles in Euclidean space can be written as
\be
	&&\mathcal{Z}(x_1,...x_N;\beta)=\nonumber\\
	&&\hspace{.2cm}\frac{1}{\mathcal{N}}\sum_P\int \mathcal{D}x_1(\tau)...\mathcal{D}x_N(\tau)e^{-\int_0^{\beta}S_E[x_1,..,x_N;\tau]\textrm{d}\tau}
\ee
where $\beta=\frac{1}{k_BT}$ and the Euclidean action is
\be
	S_E[x_1,..,x_N;\tau]=\sum_{i=1}^N\Big(\frac{m}{2}\dot{x}_1^2(\tau)+\sum_{j=1}^N\frac{V(x_i,x_j)}{2}\Big),
\ee
and only two-body interactions are taken into account. 

Now we consider the case where $K$ particles $(x_k,...x_{k+K})$ on a line move coherently, that means: $x_{k+j}(\tau)=x_k(\tau)+jd$. This hypothesis implies that we have chosen the particular permutation where $x_{i\neq k,...,k+K}(\tau)=x_i,x_k(\tau)=x_{k+1}(0), ..., x_K(\tau)=x_k(0)$. Moreover we impose that the position of all other particles is constant. 

\subsection{The ``moving string'' model}

Having selected this particular configuration, we can evaluate the potential acting on a particle lying on the line, due to all the other particles. 

\begin{figure}[!ht]
\hspace{10cm}
	\includegraphics[width=8cm]{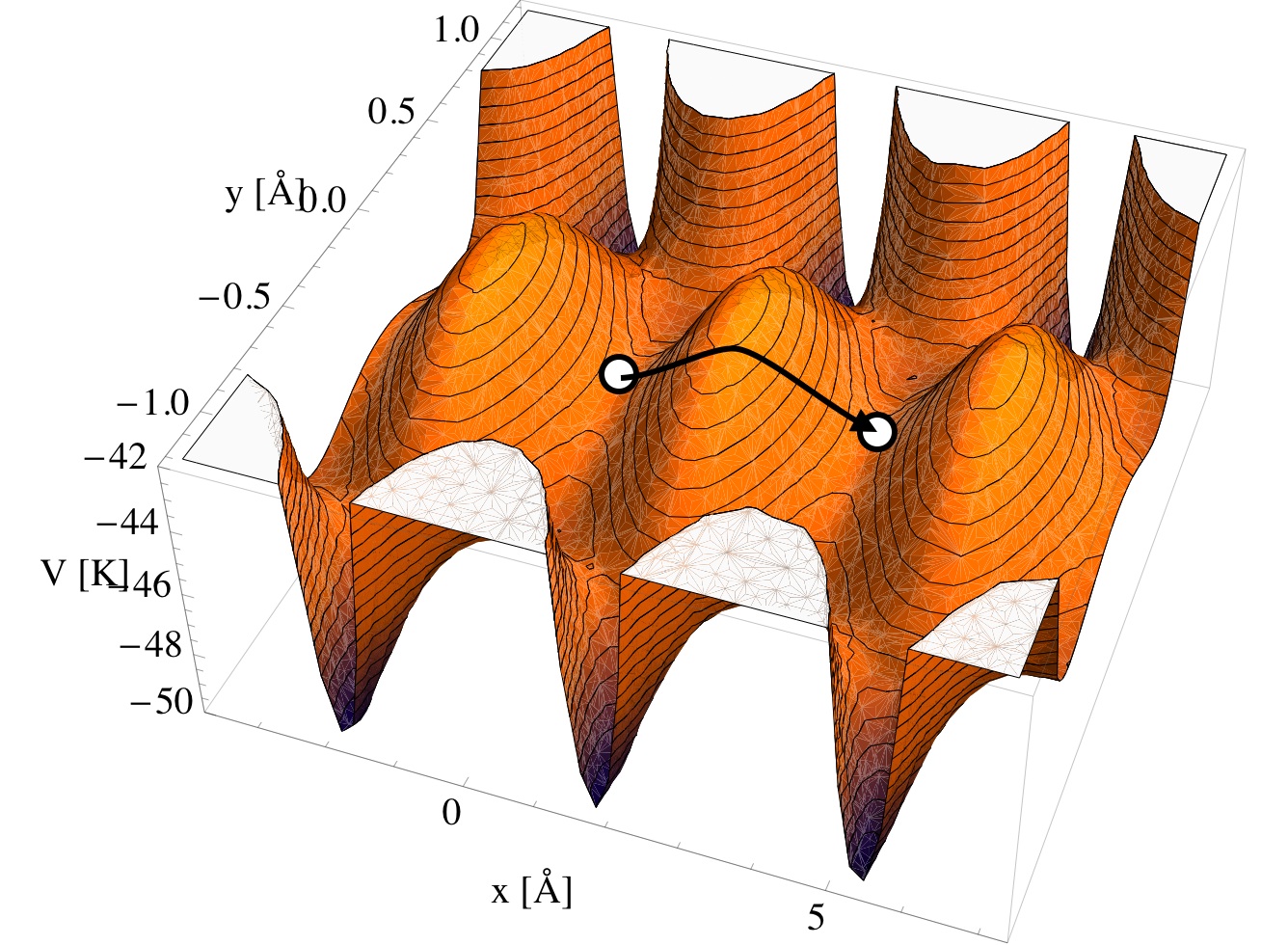}
\caption[h]{\label{fig:pot3D} Projection on the $x-y$ plane of the potential for a particle on the moving line}
\end{figure}

In Fig.~\ref{fig:pot3D} we can see the projection on the $x-y$ plane of the resulting mean field potential and one of the possible path followed by the atoms on the moving line. 

\begin{figure}[!ht]
\hspace{10cm}
	\includegraphics[width=8cm]{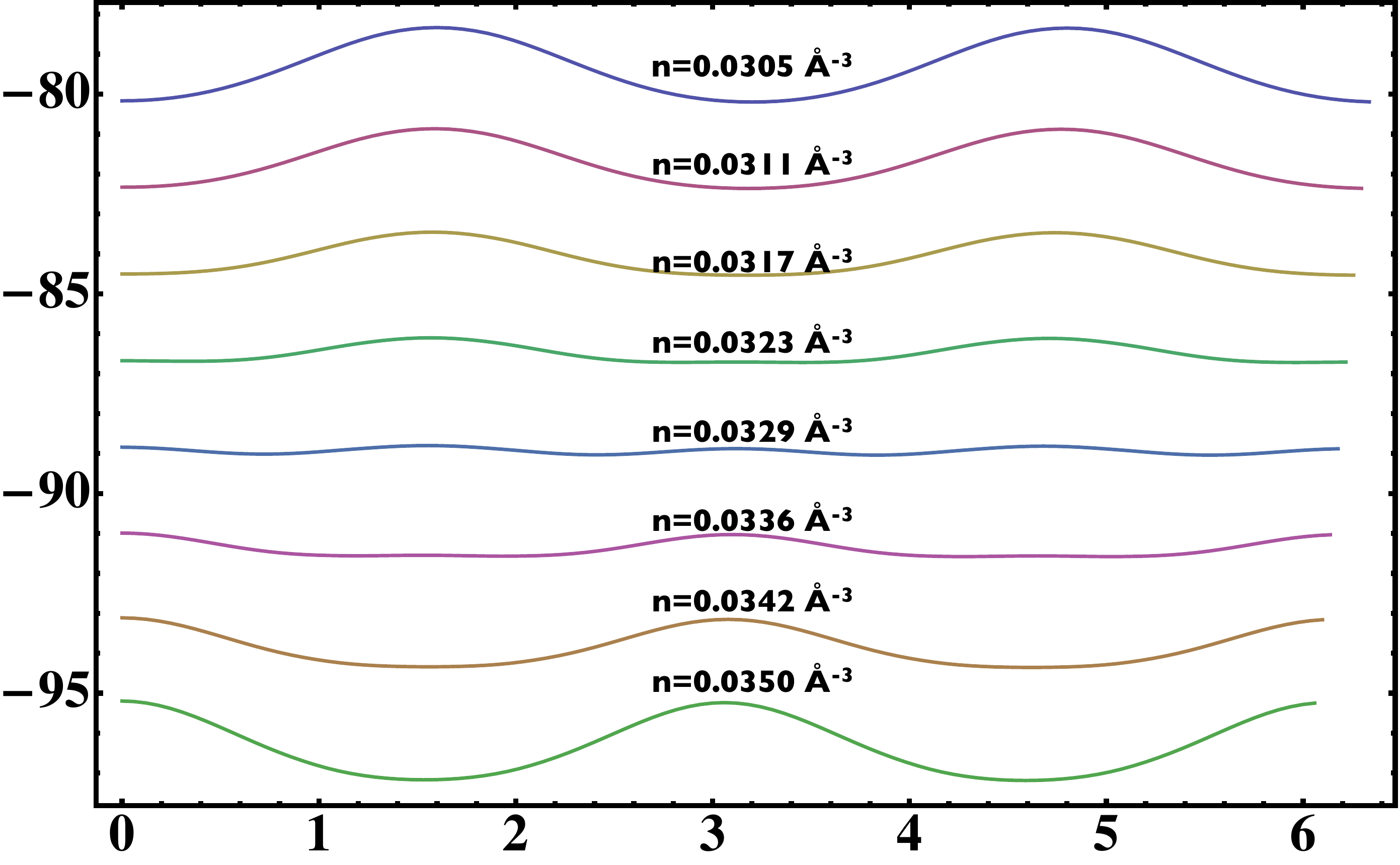}
\caption[h]{\label{fig:cubic} Density dependence of the potential when we consider the atoms on a cubic lattice}
\end{figure}

\begin{figure}[!ht]
\hspace{10cm}
	\includegraphics[width=8cm]{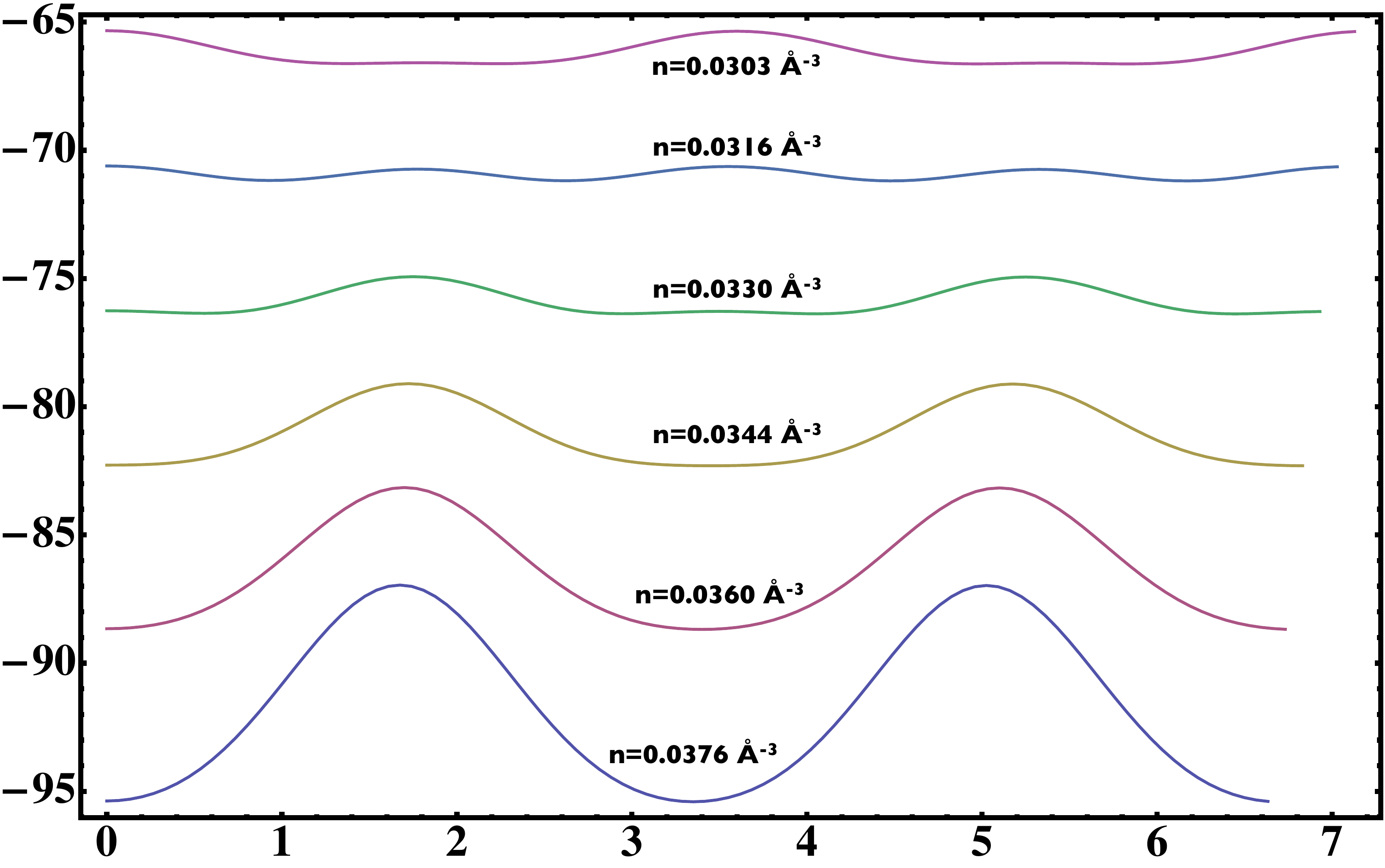}
\caption[h]{\label{fig:HCP} Density dependence of the potential when we consider the atoms on a HCP lattice}
\end{figure}

In particular we consider only the $x$-coordinate and study the dependence from the density of the mean field potential. In Fig.~\ref{fig:cubic} we can see how change the behavior of the potential considering the atoms disposed on a cubic lattice and changing the density. In Fig.~\ref{fig:HCP} we do the same for an HCP crystal. We can see that there is a range of density in both case where the origin is no longer a minimum, which means that the lattice considered is not the correct one to study the system. Therefore we need to pay attention which configuration we choose depending on the density.

\begin{figure}[!ht]
\hspace{10cm}
	\includegraphics[width=8cm]{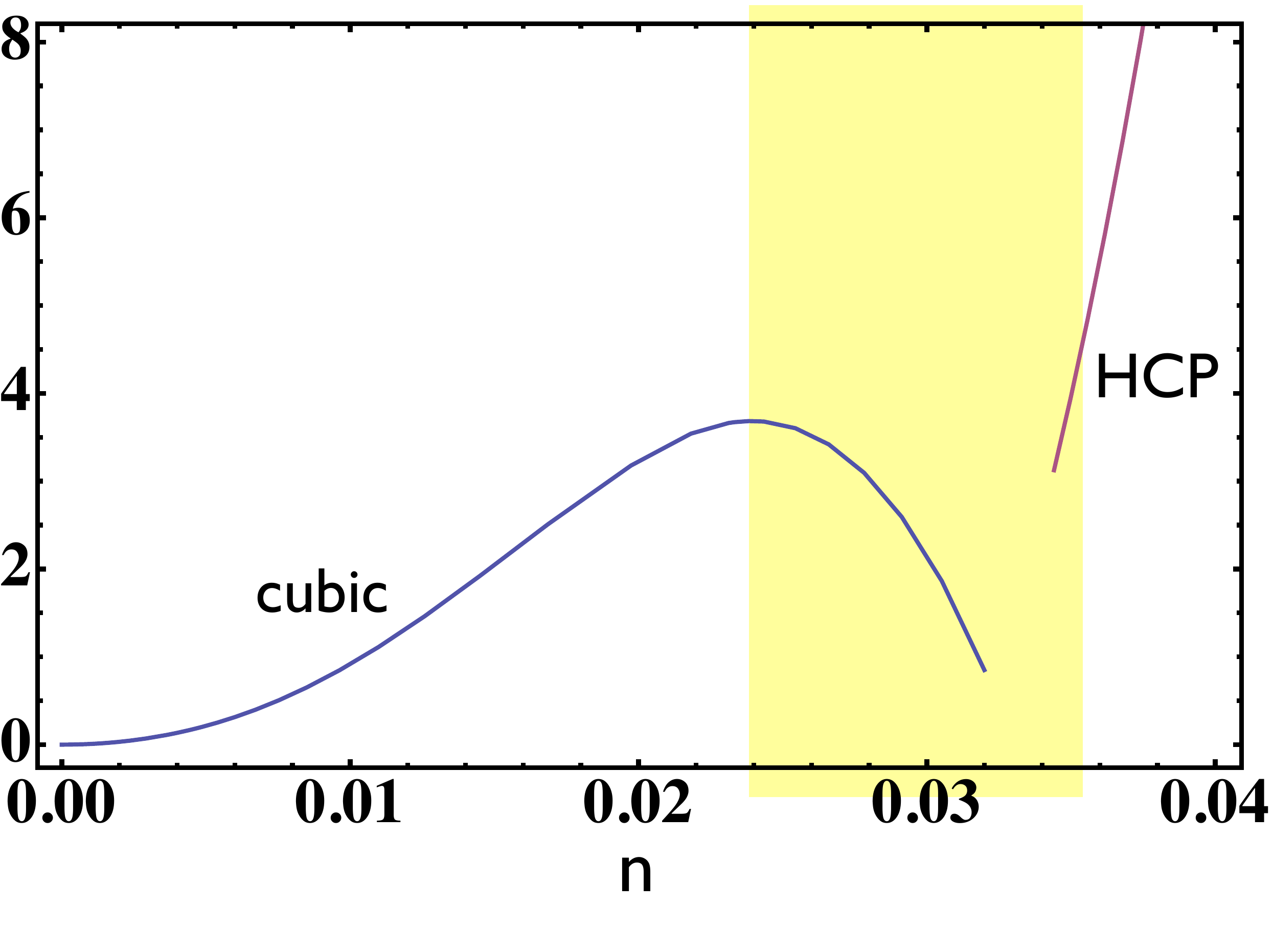}
\caption[h]{\label{fig:V0chcp} Density dependence of the amplitude of the sinusoidal potential.}
\end{figure}

Another fact related to the Aziz potential is represented in Fig.~\ref{fig:V0chcp}, where we see that the amplitude of the sinusoidal potential does not increase continuously when the density grows. We can think that such a behavior will be connected wit a wrong description of the atoms distribution in this range of densities.

This is indeed peculiar of the Aziz potential where we have both attractive and repulsive terms. If for example we consider a simple negative power potential, such a behavior not only disappears but we can find a simple relation between the amplitude $V_0$ and the density.

\begin{figure}[!ht]
\hspace{10cm}
	\includegraphics[width=8cm]{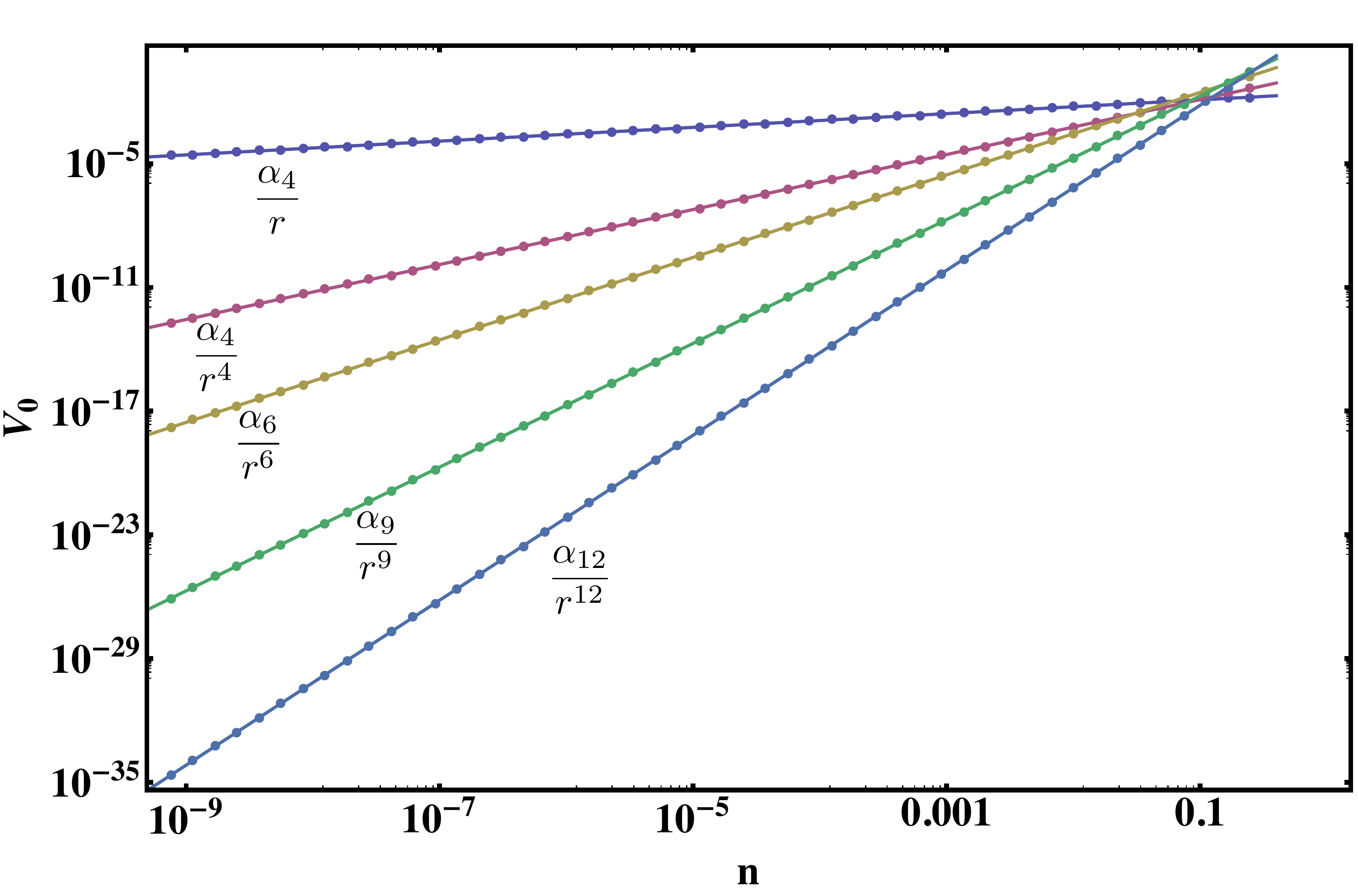}
\caption[h]{\label{fig:NPfit} Density dependence of the NP potential for five different values of $p$. The point can be successfully fitted by the function $V_0(n)=a n^{p/3}$.}
\end{figure}

Define the negative power (NP) potential 
\be
	V^{NP}=\frac{\alpha_p}{r^p}
\ee
with $\alpha_p>0$. The scattering length can be defined for this potential when $p>3$ and is given by
\be
	a=\left(\frac{2m\alpha_p/\hbar^2}{(p-2)^2}\right)^{1/(p-2)}\frac{\Gamma[(p-3)/(p-2)]}{\Gamma[(p-1)/(p-2)]},
\ee
where $\Gamma[x]$ is the Gamma function. We consider $p=1,4,6,9,12$ and choose to keep fixed the value $a=1$ for all the potentials, changing accordingly $\alpha_p$. For $p=1$ we arbitrarily fixed $\alpha_1=\alpha_4$. For this potentials we compute $V_0$ for different density, taking care that for such a NP potential the atom disposition must be a bcc crystal. The result are presented in Fig.~\ref{fig:NPfit}: the data can be fitted with extraordinary accuracy using the function:
\be
	V_0(n)=an^{p/3}
\ee

\begin{figure}[!ht]
\hspace{10cm}
	\includegraphics[width=8cm]{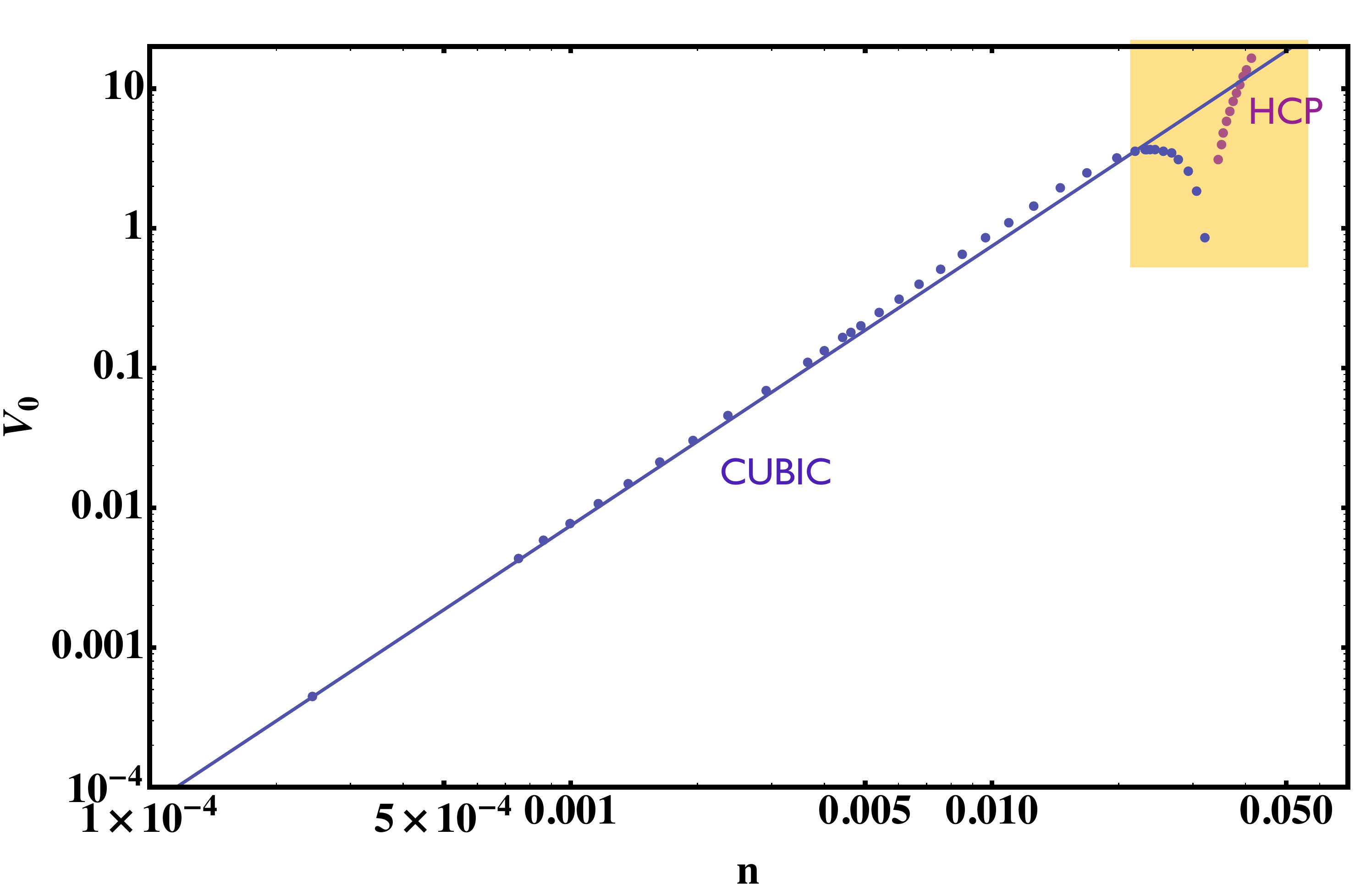}
\caption[h]{\label{fig:Afit} Density dependence of the Aziz potential fitted by the function $V_0(n)=a n^{6/3}$.}
\end{figure}

We can try to use the same function also to fit the sinusoidal amplitude obtained by the Aziz potential, in this case we can use $V_0(n)=a n^{6/3}$ where $p=6$ is the dominating term for large distances that appear in the Aziz. The result is in Fig.~\ref{fig:Afit}, we can see that the fit is quite good until, as expected, we enter in the yellow band described before.

Nevertheless we decided to use this simple scaling relation for our calculations.

Due to the particular configuration we have chosen, the partition function for the Bose system can be rewritten as
\be
	&&\mathcal{Z}(x_1,...x_N;\beta)=\nonumber\\\nonumber
	&&\frac{1}{\mathcal{N}}\Big[\int\mathcal{D}x_k(\tau)e^{-\int_0^{\beta}\Big(\frac{m}{2}\dot{x}_k^2(\tau)+C_0(x)\Big)\textrm{d}\tau}\Big]^{(N-K)}\cdot\nonumber\\
	&&\hspace{.3cm}\Big[\int\mathcal{D}x_n(\tau)e^{-\int_0^{\beta}\Big(\frac{m}{2}\dot{x}_n^2(\tau)+V_0(x)\Big)\textrm{d}\tau}\Big]^K
\ee
where $C_0(x)$ is the potential acting on the particle at rest and which is exactly the potential that also appears in the Boltzmann partition function.

For this reason the ratio between the Bose ($\mathcal{Z}$) and the Boltzmann ($\mathcal{Z}_B$) partition function reduces to
\be\label{eq:rBB}
	\frac{\mathcal{Z}}{\mathcal{Z}_B}=\Bigg[\frac{\int\mathcal{D}x_k(\tau)e^{-\int_0^{\beta}\Big(\frac{m}{2}\dot{x}_k^2(\tau)+V_0(x)\Big)\textrm{d}\tau}}{\int\mathcal{D}x_k(\tau)e^{-\int_0^{\beta}\Big(\frac{m}{2}\dot{x}_k^2(\tau)+C_0(x)\Big)\textrm{d}\tau}}\Bigg]^K
\ee
and the Feynman parameter is given by 
\be\label{eq:yfN}
	y_F=\frac{\int\mathcal{D}x_k(\tau)e^{-\int_0^{\beta}\Big(\frac{m}{2}\dot{x}_k^2(\tau)+V_0(x)\Big)\textrm{d}\tau}}{\int\mathcal{D}x_k(\tau)e^{-\int_0^{\beta}\Big(\frac{m}{2}\dot{x}_k^2(\tau)+C_0(x)\Big)\textrm{d}\tau}}
\ee
\subsection{The semiclassical calculation}
The simplest case is that of noninteracting particles, for which the
 optimal path (``Feynman's instanton")
 is just the straight line starting at the initial position of a
particle and
ending at the previous position of its neighbor
(as indicated by  lines with arrows in Fig.~\ref{fig_latt}).  Since velocity
is constant, the extra action per particle
 needed for its jump
is nothing but just the kinetic energy times the Matsubara time  $\beta=\hbar/T$ available for the interchange, with an obvious velocity on the segment being $d/\beta$ 
\be 
	S_{ideal} = {m \over 2} ({d\over  \beta})^2 \beta 
\ee
per particle, for each of the diagrams. Let us see what is this action at the Bose-Einstein condensation (BEC)line, for the ideal gas. Using 
\be T_c=3.31 \hbar^2 n^{2/3} /m \ee
and relating the distance of the jump to the density by $d=\kappa n^{-1/3}$, with some coefficient $\kappa=O(1)$ 
%
 one finds that the critical 
action for the ``Feynman instanton'' is  
\be 
S_{BEC}/\hbar={3.31\kappa^2 \over 2} =\kappa^2 1.655 
\ee
The precise value depends on the numerical constant $\kappa$, which in turn depends how exactly particles are correlated in space.
It is unclear what should it be for ideal gas, but for strongly coupled systems we are mostly interested in it
is uniquely determined by the type of local crystal structure developing, e.g. it is exactly $\kappa=1$ for cubic crystal.
We will use below this value, although there is some uncertainty here for weakly coupled systems.

When density is too low or $T$ is too high, so that $S>S_{BEC}$,
 the sum over the polygons is exponentially convergent and thus the
 gas remains in normal phase.

Now we switch on the potential and impose our atoms to be displaced on
 a cubic lattice.
As a first analytic approach to the problem
 we look for the semiclassical tunneling path -- known as periodic
 instanton, or caloron. It corresponds to solution of classical
 equation of motion in Euclidean time, now including the
potential. All cases we are interested in this periodic
potential on various lattices and forces  happens to be well
described its first harmonics, the sinusoidal potential:
\be
	V(x)=V_0\sin^2(x\pi/d)
\ee
which interpolates between the minima of $V$
\be
	x_{cl}(0)&=&0\nonumber\\
	x_{cl}(\beta)&=&d\nonumber
\ee
Introducing conserved Euclidean energy $E_E$
\be
	E_E=\frac{m}{2}\dot{x}^2-V(x) &\rightarrow& \dot{x}=\sqrt{\frac{2}{m}(V(x)+E_E)}
\ee
after separation of variable we has the solution
\be
	\frac{\textrm{d}x}{\sqrt{\frac{2}{m}(E_E+V(x))}}=\textrm{d}\tau
\ee
which can be easily integrated
\be\label{eq:tau}
	&&\sqrt{\frac{m}{2}}\int_{x_{cl}(0)}^{x_{cl}(\tau)}\frac{\textrm{d}x}{\sqrt{E_E+V(x)}}=\nonumber\\
	&&\sqrt{\frac{m}{2}}d \frac{\textrm{F}[x\pi/d,-V_0/E_E]}{\pi\sqrt{E_E}}|_{x_{cl}(0)}^{x_{cl}(\tau)}=\tau-\tau_0
\ee
where F is the elliptic integral of the first kind. Imposing $x_{cl}(\tau_0=0)=0$ we have
\be
	x_{cl}(\tau)=\frac{d}{\pi}\textrm{JA}\Big[\sqrt{\frac{2E_E}{m}}\frac{\pi}{d}\tau,-\frac{V_0}{E_E}\Big]
\ee
with JA the Jacobi amplitude for elliptic functions.
 Particular solution for $E_E=0$ is called the instanton,
it corresponds to the zero temperature ( or  $\beta\rightarrow\infty$)
limit. In our setting it is not interesting since we only
approach the critical point from above and thus $T>T_c$.

For arbitrary temperature the duration of the Matsubara time $\beta$
is prescribed: thus 
from Eq.~\ref{eq:tau} we obtain the equation to fix the
energy. One can also
 read it immediately
as the critical temperature as a function of the parameter of our system ($d$, $V_0$ and $E_E$)
\be\label{eq:tc}
	&&T_c[d,V_0,E_E]=(\sqrt{\frac{m}{2E_E}}\frac{d}{\pi}\textrm{F}[\pi,-V_0/E_E])^{-1}\nonumber\\
	&&=\sqrt{\frac{m}{2E_E}}\frac{d}{\pi}2\textrm{K}[-V_0/E_E])^{-1}.
\ee
provided the energy is substituted from the Feynman condition,
which say that
the Euclidean action
\be\label{eq:se}
	&&S_E[d,V_0,E_E]=\int_0^{\beta}\textrm{d}\tau\Big(\frac{m}{2}\dot{x}_{cl}^2+V(x))\nonumber\\
	&&=\frac{d}{\pi}\sqrt{\frac{mE_E}{2}}2\Big(2\textrm{E}\Big[-\frac{V_0}{E_E}\Big]-\textrm{K}\Big[-\frac{V_0}{E_E}\Big]\Big),
\ee
is equal to the Feynman critical value.

\begin{figure}[!ht]
\hspace{10cm}
	\includegraphics[width=5cm]{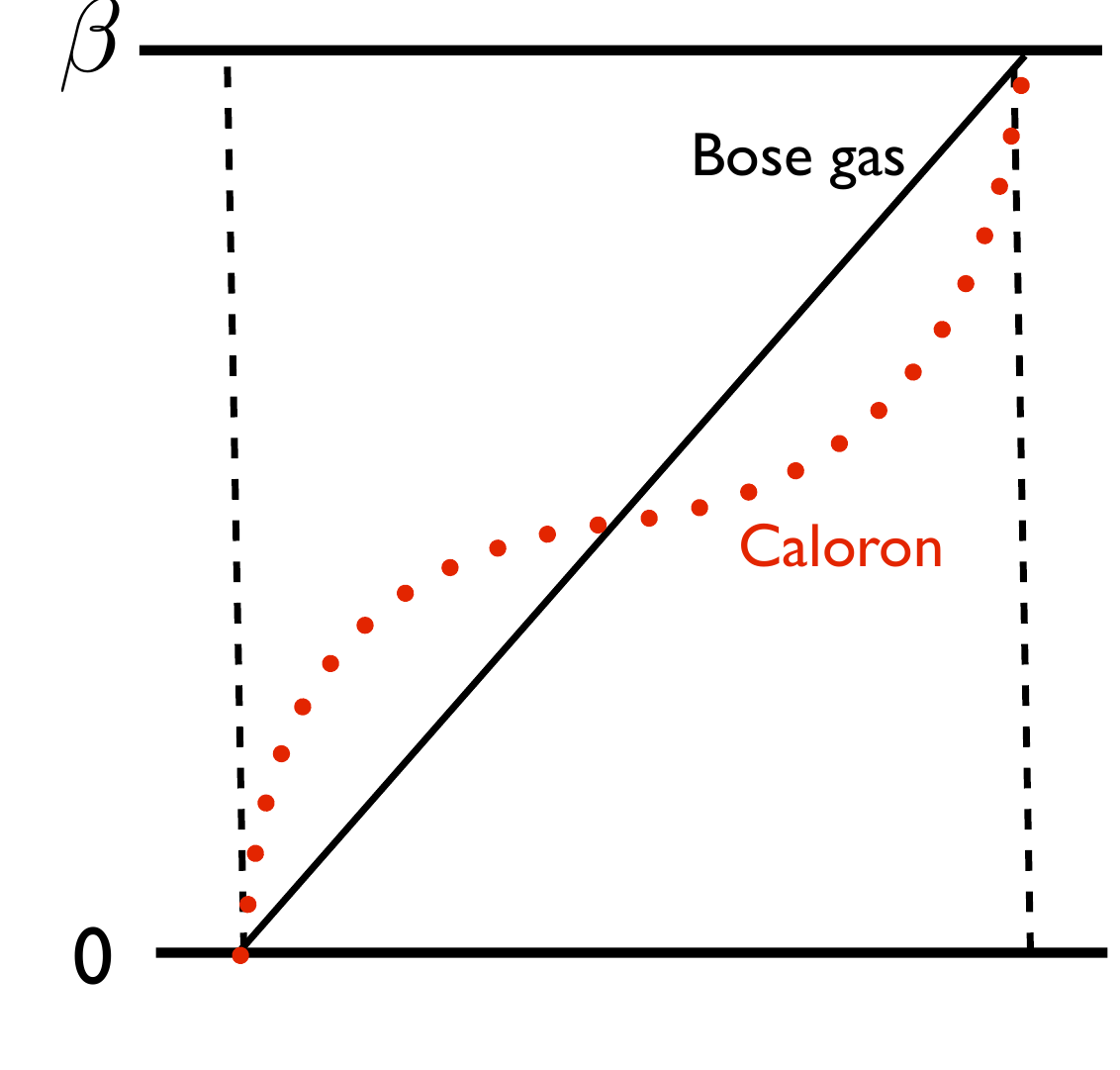}
\caption[h]{\label{fig:path} Instanton trajectory we have in the presence of a potential barrier, compared to the free Bose gas case.}
\end{figure}

In the instanton limit $E_E=0$ we have for the action
\be S_0=\frac{ 2 d\sqrt{2m V_0} }{ \pi}  \label{eqn_instanton} \ee
And the corresponding trajectory is given by
\be
	x(\tau)=\frac{2}{\pi}d\ \textrm{arccot}\left[\exp\left(-\sqrt{\frac{m}{2}}\frac{\pi V_0}{d}\tau\right)\right].
\ee
In Fig.~\ref{fig:path} we can see the difference between the instanton trajectory and the simple straight line corresponding to the case of a free Bose gas.
\vspace{0.5cm}

\begin{figure}
\hspace{10cm}
	\includegraphics[width=8cm]{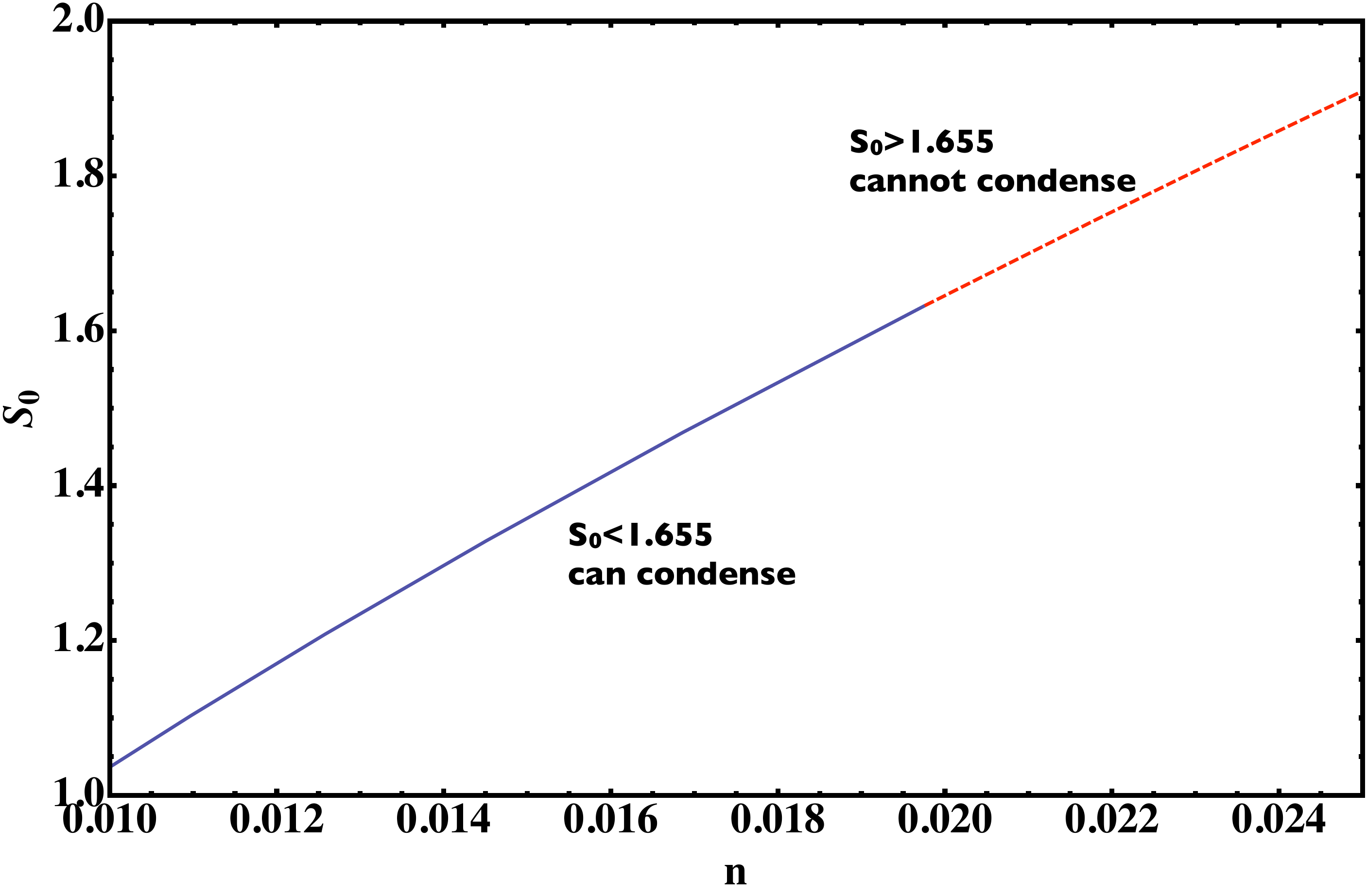}
	\includegraphics[width=8cm]{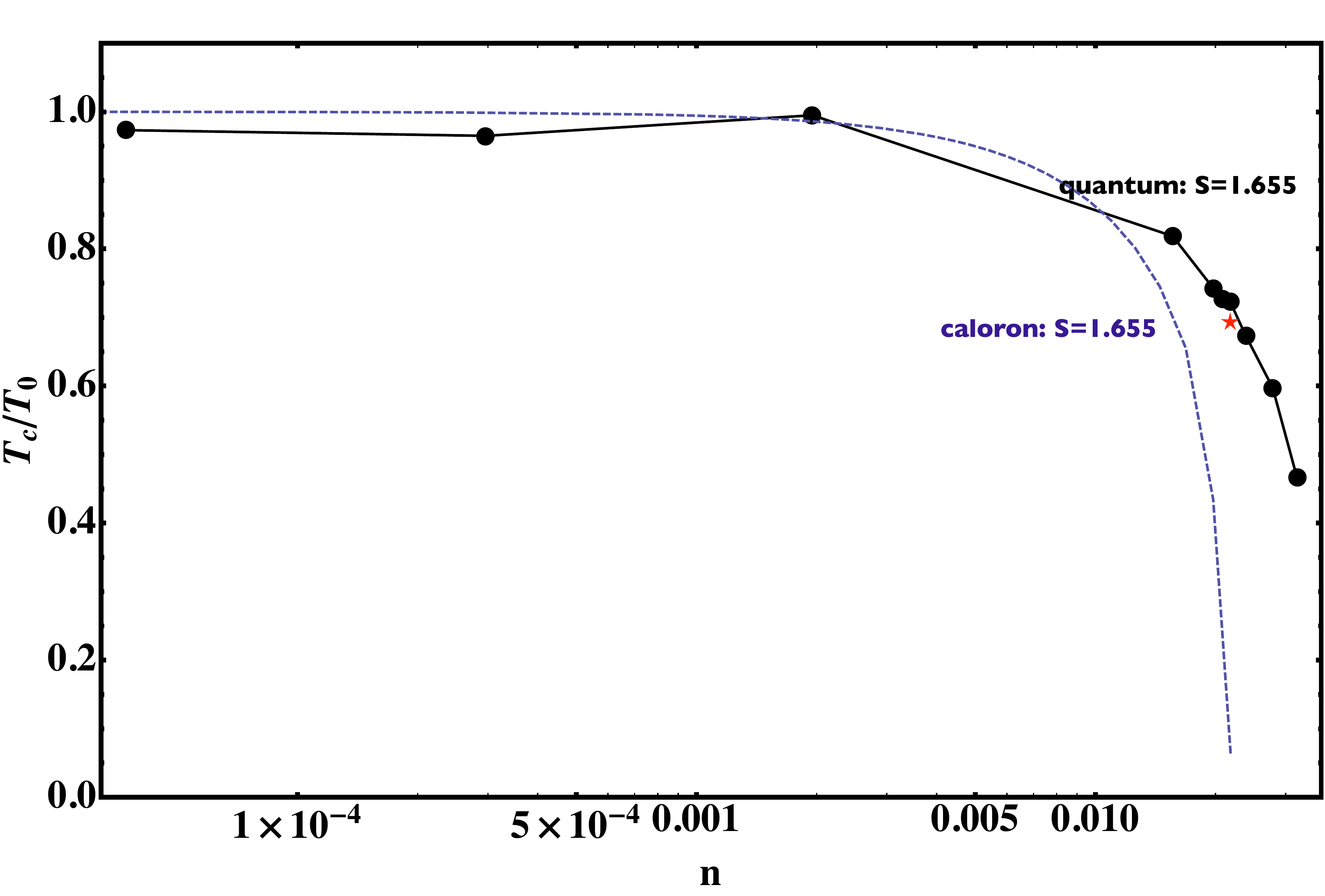}
\caption[h]{\label{fig:S0inst}(a) The action for the tunneling (caloron) solution
  as a function of the density. When the action is larger than the
  critical value $S=1.655$ (dashed line) the system cannot be Bose-Einstein
 condensed phase.
\\ (b) The critical temperature obtain from the caloron/instanton
solution (dashed line) and from the 1-d PIMC simulation (points). The red star
  shows the physical location of the lambda point.}
\end{figure}

 More detailed comparison 
between real He and caloron/instanton approximation is shown in
 Fig.~\ref{fig:S0inst}(b). While it is qualitatively correct,
it is not quite accurate. This is by no means surprising: it should be accurate provided
the critical action for BEC would happen to be much larger than 1, while
 unfortunately it is only 1.65 or so. Therefore the semiclassical result should be used for qualitative comparison only.


One of those is that as the matter is further compressed by extra pressure, 
its density and the amplitude of the potential $V_0(n)$ grow. Respectively the action required for
  tunneling grows, see Fig.~\ref{fig:S0inst}(a), and when  the action
 gets too large $S_0>S_c$, Feynman's condition could not be
 fulfilled. Therefore, a sufficiently compressed ${}^4$He $cannot$ support 
a supercurrent of the ${}^4$He atoms.
 
 (For clarity: we do not make any statements here about possible ``supersolid" 
 behavior of the solid ${}^4$He induced by supercurrent of some defects/dislocations imposed on it. All we are saying is that the ${}^4$He atoms themselves
 do not create such supercurrent:  this statement is however not  new and it has been  verified in dedicated numerical Monte-Carlo studies. The only new 
 element in our statement is that it follows from Feynman's universal action.)

 One may improve the semiclassical expressions for the tunneling action in one or two 
  loop  quantum corrections. For example with the one-loop accuracy 
the  Feynman parameter can be written as
\be
	y_F=\frac{e^{-\frac{d}{\pi}\sqrt{\frac{mE_E}{2}}2\Big(2\textrm{E}\Big[-\frac{V_0}{E_E}\Big]-\textrm{K}\Big[-\frac{V_0}{E_E}\Big]\Big)}}{\Big(\textrm{det}\hat{F}[x_{cl}]\Big)^{1/2}}.
\ee
Expressions for two loop are a bit more involved, see Ref.\cite{Wohler:1994pg}.
However we have not calculated it analytically, using numerical
path integral instead.

\subsection{The 1-d quantum path integral}
As we have already emphasized above, since the action is not large, one cannot expect the semiclassical theory be really accurate. Fortunately quantum fluctuation around the classical trajectory can be taken into account
by numerical evaluation of the r.h.s. of (\ref{eq:yfN}) can be simply performed using PIMC code. Note that compared to PIMC simulation of manybody these calculations are very cheap in terms of computational power because we consider a one dimensional system with only one particle.

\begin{figure}
\hspace{10cm}
	\includegraphics[width=8cm]{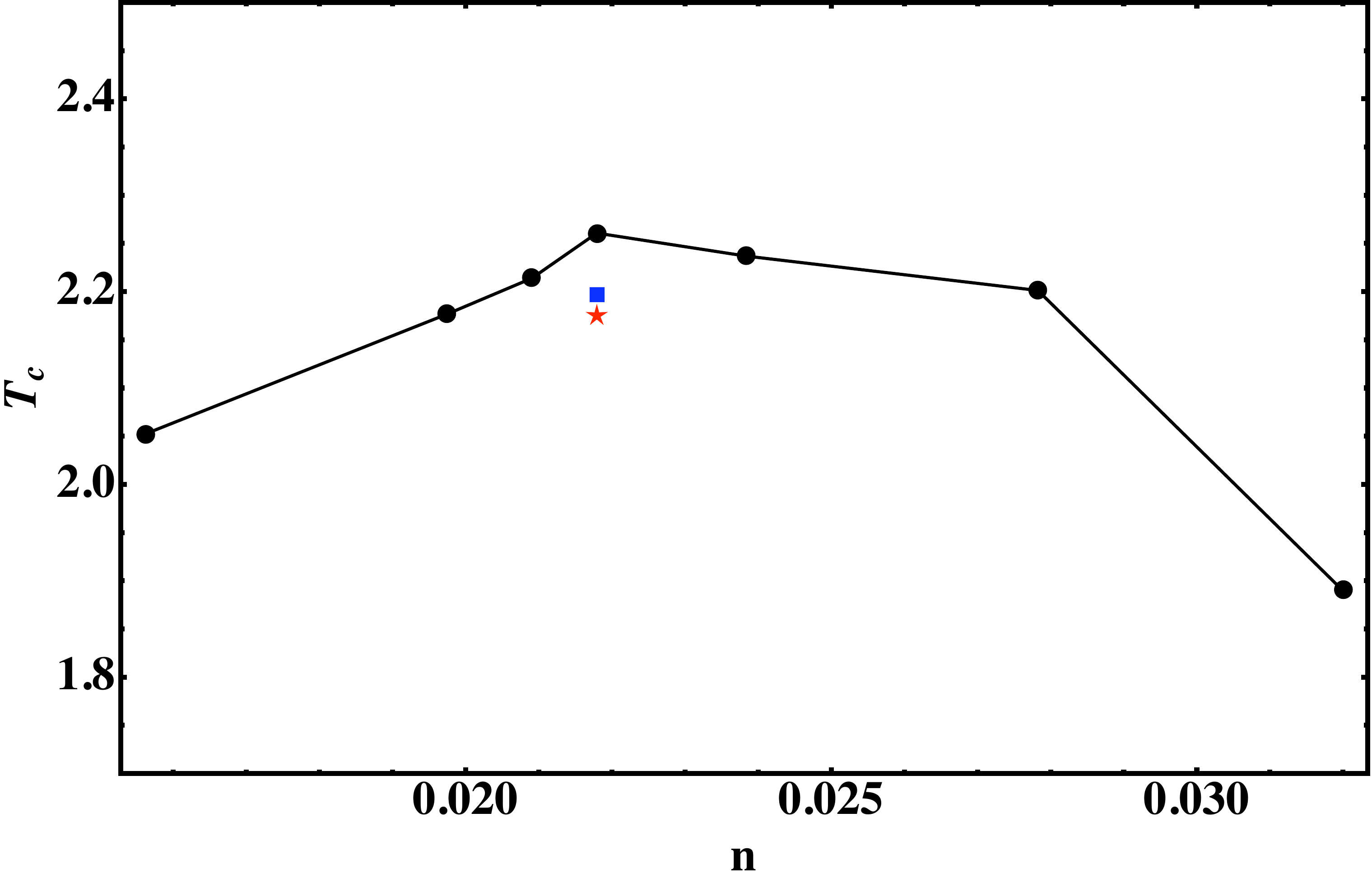}
\caption[h]{\label{fig:tcc} Density dependence of the critical temperature. Black points are obtained using the exponential dependence of $V_0(n)$.The red star is the physical result for ${}^4$He. The blue square is obtained using the $V_0$ as obtained directly from the Aziz potential.}
\end{figure}

In order to find the critical temperature as a function of the density we do the ansatz that the value of the Feynman parameter of our system is more or less the same as for the free Bose gas. Because we know that for the free Bose gas $S_F=-\log(y_F)=1.655$ independent from the density, we simply computed $S$ for different temperature at fixed density and when $S(T,n)=1.655$ we simply called that critical temperature. in Fig.~\ref{fig:tcc} we plotted the results obtained in the case of the cubic lattice. We can see the good agreement between our calculation and the physical critical point. Also when we use the formula  $V_0(n)=a n^{6/3}$ for the prediction of the amplitude, our prediction is only four percent larger than the physical value. 

Is also interesting to notice that the physical point is the point where the critical temperature is larger, in this sense is looks like natural that condensation happens in nature exactly at that point.

Now we plot the ratio $T_c/T_0$ (Fig.~\ref{fig:S0inst}~(b) black points), where $T_0$ is the critical temperature for the free Bose gas. In the limit of vanishing density we recover the Einstein's critical temperature as it should be. In the region of small density there is universal theory using small parameter $a^3n$, with $a$ being the scattering length.
As shown e.g. in Ref.\cite{Holzmann:2001zz},
  there is a nontrivial  increase of $T_c$ by several percents there, in agreement with numerical studied.
Our model is obviously too crude to  reproduce it.

\begin{figure}
\hspace{10cm}
	\includegraphics[width=8cm]{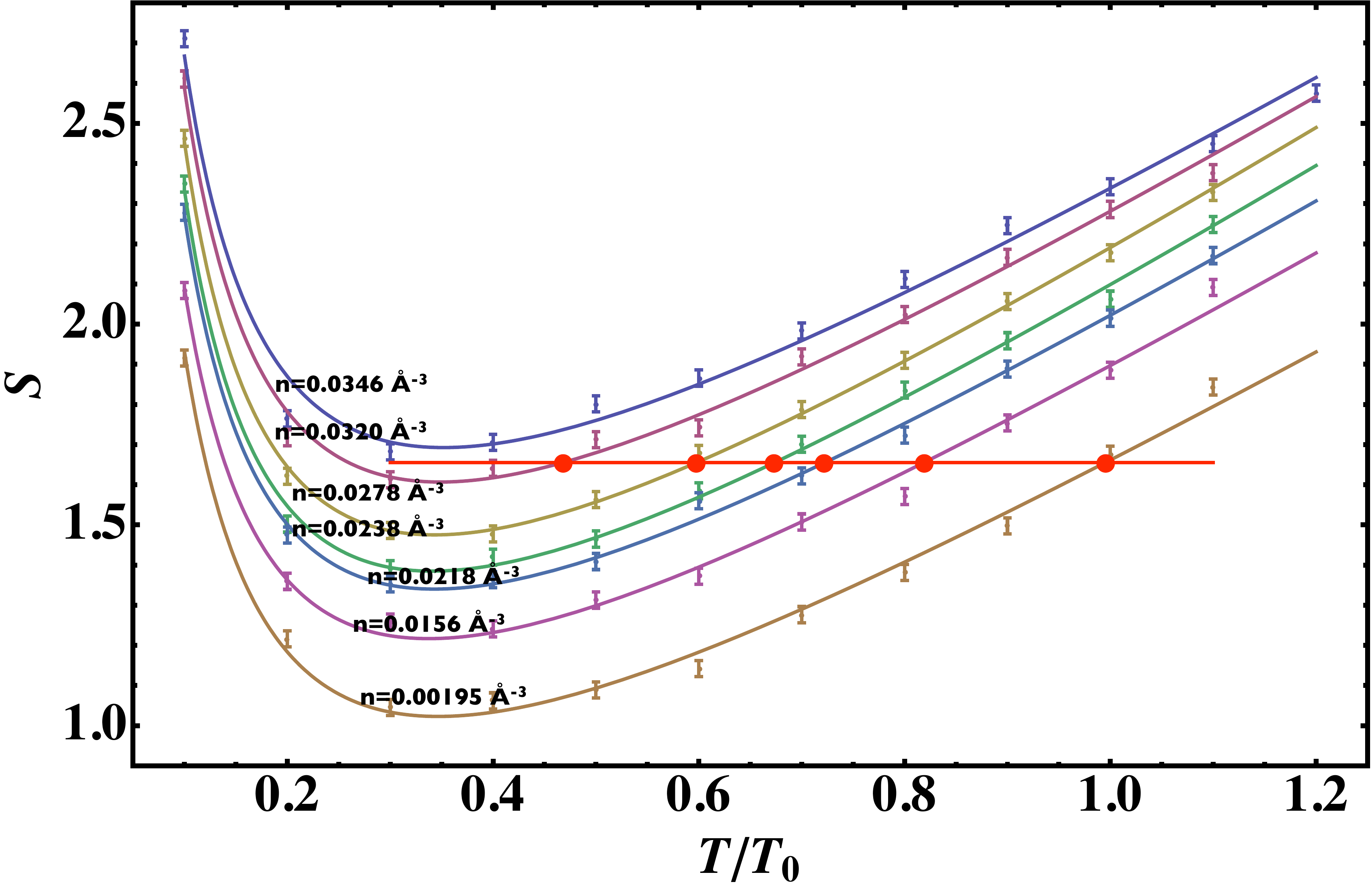}
\caption[h]{\label{fig:actsin} Action as a function of the temperature for different densities.}
\end{figure}

 We are mostly interested in the opposite limit of high density: here the considered ratio decreases until a certain point. For larger density we simply cannot have condensation in the sense that we cannot find any temperature where $S(T)=1.655$. This can be seen more clearly looking at Fig.~\ref{fig:actsin}. Starting from around $n\simeq0.033\ \AA^{-3}$ we cannot find any critical temperature different from zero. As a consequence we have a jump from a minimum of $T_c\simeq1.7$ K to $T_c=0$ around that density.

\begin{figure}
\hspace{10cm}
	\includegraphics[width=8cm]{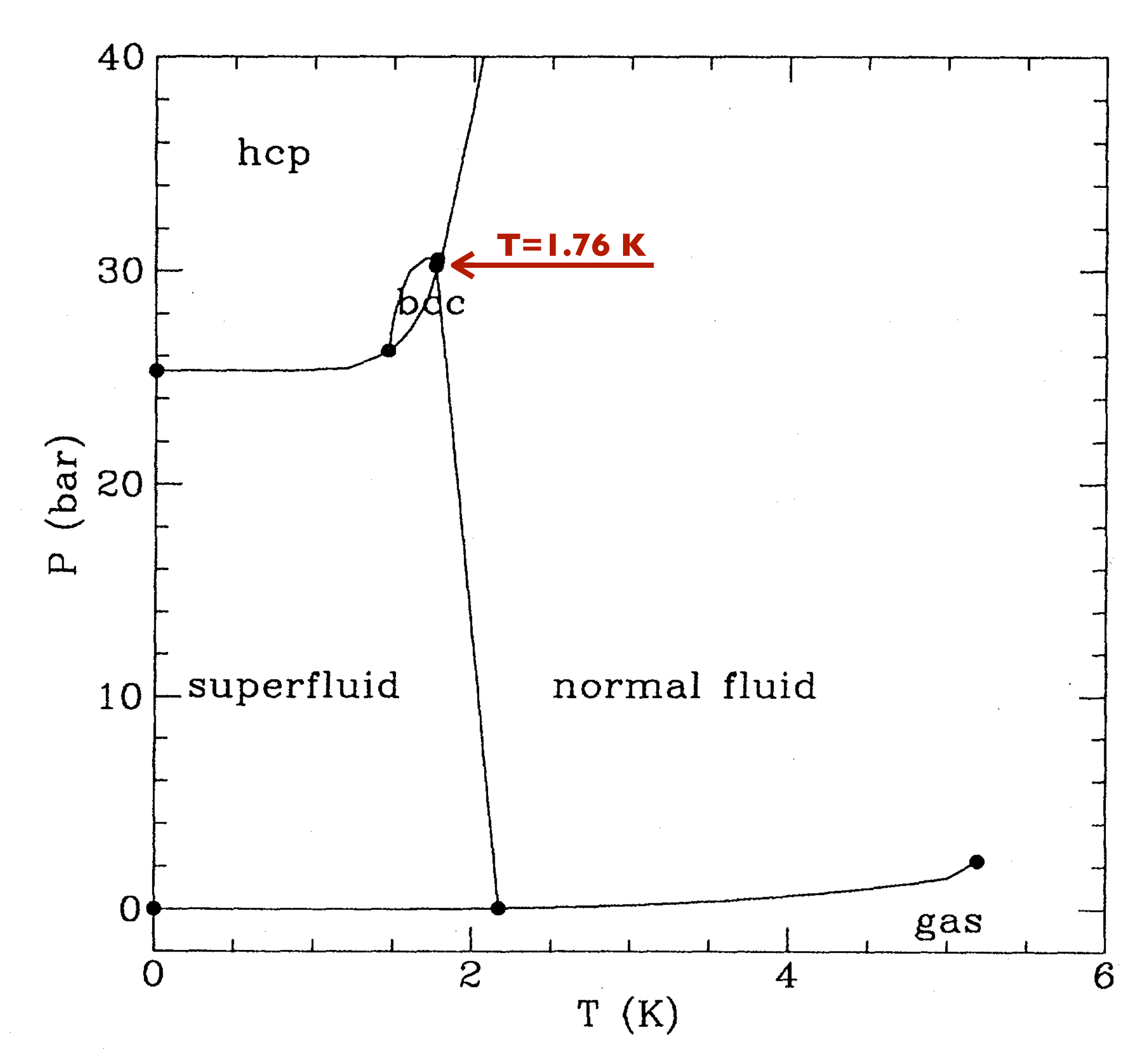}
\caption[h]{\label{fig:Hepd} ${}^4$He phase diagram}
\end{figure}

Looking at the ${}^4$He phase diagram (Fig.~\ref{fig:Hepd}) is also interesting to notice that the $\lambda$-line starts at $P=0$, $T=2.17$ and ends at some pressure with $T=1.76$ very close to our estimate for the minimal critical temperature. Obviously the density where we observe this behavior is completely different, nevertheless is interesting to notice that the two minimal critical temperature coincide. It seems to suggest that it is not possible to have a supersolid phase of Helium because, as soon as we enter in the non-liquid phase, the critical temperature jumps from $T_c=1.76$ to zero.

\section{Confinement as Bose-Einstein condensation of monopoles}\label{sec:mon}

First of all, let us explain why we think Feynman criterion based on particle paths and clusters is especially 
well suited for this task. One reason is that lattice simulations are able to follow monopole's paths, and thus identify
``single" ones, with individually periodic paths, as well as those belonging to k-clusters. Therefore, one can see
divergence of the cluster expansion at $T_c$ directly, without any calculations. 

The second reason is that excited matter we discuss has no nonzero quantum numbers, 
it is just an  ``excited vacuum'' produced in high energy collisions.
Thus there is no any nonzero conserved charge to which the corresponding chemical potential can be coupled. 
 Neutrality leads to  equal number of electrically charged  quarks and antiquarks, as well as equal number of 
 monopoles and anti-monopoles. Thus unfortunately one cannot introduce chemical potential and use the usual
 reasoning related to its crossing the lowest level.
 
 And yet , since the density and mass of such monopoles depend on $T$,  only when the Feynman criterion is satisfied their BEC may happen.
 In this section we estimate what mass and coupling constant the monopole should have in order
 to condense. We do so in two steps, first for noninteracting monopoles, including relativistic
 action, and then for interacting ones using nonrelativistic formulae derived above.
 Again following Feynman,  we approach the problem from the high-T phase -- the QGP -- in which monopoles are not  Bose-condensed,
calculate the action needed for a ``jump" to the site of the identical neighbor, and compare it to the universal value $S_c$.
 
 We will only consider pure gauge theories (no quarks) and ignore gluon quasipaticles. The standard units used in this field are based on $\hbar=c=1$, and thus only one unit (length in fm or energy in GeV, so fm*GeV=$0.1973$) needs to be defined. It is  standard in lattice works to calculate the string tension $\sigma$ at zero T and fixed its value to be the same as in real QCD, namely $\sqrt{\sigma}=0.42\, GeV $, which we will follow. The critical temperature in such units is $T_c\approx 0.27\, GeV$.
Lattice $T_c$ for pure $SU(3)$ gauge theory is in such units about 270 MeV.

\subsection{Free monopoles}

The work \cite{D'Alessandro:2007su} (pure gauge SU(2) theory) has provided measurements for monopole (plus antimonopole) density~$n$ at $T=(1-12)T_c$. In this theory Higgsing leaves only one massless U(1), and thus there is only one kind of monopole. For orientation, near $T_c$ they find $n/T^3\approx 0.3$. This the distance between identical monopoles is 
\be  a=(n/2)^{-1/3}=1/(0.53 *0.27 GeV)\approx 1.4\ \textrm{fm} \ee
(where 2 is because we only need density of one charge).

We now estimate the action of the monopole which jumps to the position of another identical monopole at distance $a$ away, during 
 Euclidean time duration equal to the critical Matsubara time $\beta=\hbar/T_c=1/(0.27\, \textrm{GeV})=0.73\, \textrm{fm}$.
In the Feynman approximation -- when only kinetic part of the action is included -- the (Euclidean) velocity on the optimal path is constant $v=a/\beta$. Putting numbers reveals two complications: (i) $v>1$ and (ii) relativistic monopoles require relativistic form for Euclidean action. Since we speak about tunneling, having imaginary action and velocity 
above that of light is in fact appropriate:  no negative roots appear. The Euclidean action at the BEC $T_c$ point should be 
\be\label{eq:se_rel} S_E=m\int ds= m \beta \sqrt{1+a^2/\beta^2} = S_c \ee 
Putting numbers into the square root one finds
\be m=T_c*S_c/\sqrt{5}\approx  200\, \textrm{MeV}\ee
From existing lattice data on the mass, estimated from the paths themselves by D'Elia (private communication) one finds rapid decrease
of the monopole mass as $T\rightarrow T_c$ to similar ball park, but the accuracy of the data is not yet sufficient to tell its numerical value at $T_c$ yet.
And before this comparison is made, we should move on into much more involved estimate for strongly interacting monopole plasma.

\subsection{Strongly interacting monopoles}
Color monopoles interact via Coulomb-like magnetic forces related to their charges. We remind the reader that
``Higgsing" is assumed to be due to nonzero background field $<A_0>$, which lives certain
 Abelian subgroups of the gauge
field  unbroken, that is massless. For the SU(2)
color group there is only one diagonal generator $\tau^3$ which is left massless, so there is only one $U(1)$ charge and  
 thus we can use the language of electrodynamics, simply calling monopoles to be positive and negative
charges. (This is strictly speaking not true  for physical $SU(3)$ color, which has  two unbroken
Abelian fields, proportional to $\lambda^3,\lambda^8$ Gell-Mann
matrices, with two families of monopoles.) One more comment is that in general forces include also Higgs exchange, which in principle
can also be long range if Higgs is massless. However the data about monopole correlations \cite{D'Alessandro:2007su} seem to show only pure Coulomb-like forces.

Such charges obviously prefer alternating 
 cubic lattice. Furthermore, classical Molecular Dynamics which
reproduces their static correlation functions observed
on the lattice (see discussion and references in
 \cite{Shuryak:2008eq}) complement the Coulomb potential
by the repulsive core, to ensure classical stability of the lattice.
The potential is thus written as 
\be V_{ij}(r) ={g_m^2 \over 4\pi} [ {(-)^{Q_i+Q_j} \over r} + {1 \over b r^b}]\ee
where charges $Q_i,Q_j=\pm 1$ and additional dimensionless parameter $b$ is the ``core power".
It is usually selected to be large, to produce small corrections
except close to the origin: we use the conventional b=9.
The forces are balanced at distance 1, which defines the
 normalization.

\begin{figure}[!ht]
\vspace{0.75cm}
\includegraphics[width=7 cm]{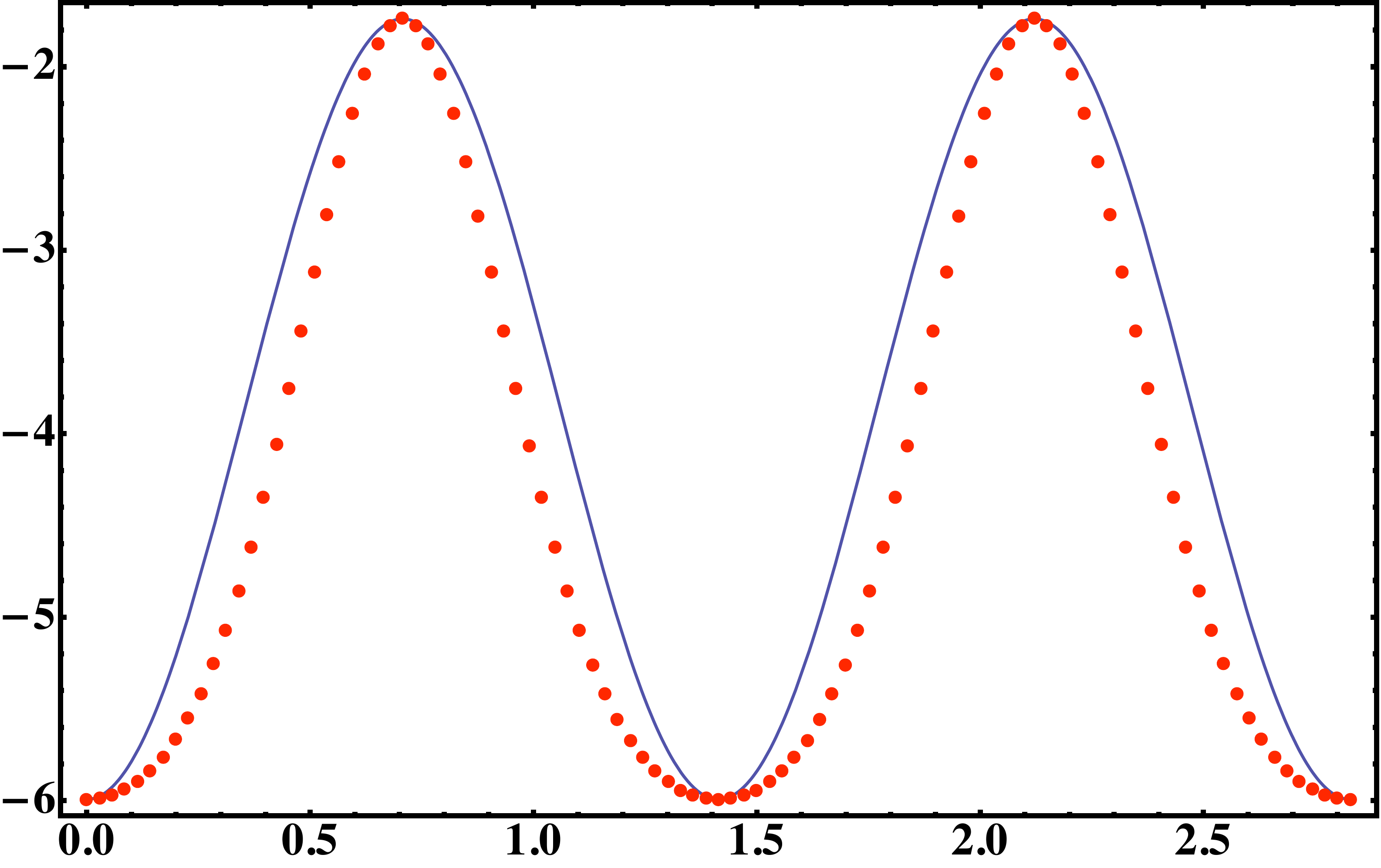}
\caption[h]{\label{fig_diag} The effective potential for a diagonal
moving line of charges (points) is compared to a fitted sinusoidal
potential (curve).
}
\end{figure}

Near $T_c$ the condensate is still small fraction of all particles,
in spite of divergent (or highly peaked) specific heat:
thus we may think that only particles of one kind (e.g.
only positive ones) are moving  along the line.
This prompts us to take {\em one diagonal} of the cubic lattice
(rather than a line parallel to axes) to be jumping the
distance $\sqrt{2}$, while keeping all other positive and 
 $all$ negative ones  stationary.
The potential corresponding to such forces
 was calculated and plotted in Fig.~\ref{fig_diag}
(points). As usual, it is well represented by  the sinusoidal potential  \be\label{eq:effpotCoul} V={g_m^2 \over 4\pi d} [C+V_0\sin^2({\pi x \over \sqrt{2}d})]\ee 
and therefor one can use the results obtained for ${}^4$He in Section~\ref{sec:feynm} (note that because of the diagonal supercurrent we have substituted $d\rightarrow \sqrt{2}d$ into it). In particular, we will use caloron amplitude~(\ref{eq:se}). This formula was derived starting from a non-relativistic Lagrangian, while we have shown that in this regime monopoles move relativistically. In order to give an estimate of the relativistic correction we can compare the monopole mass predicted using Eq.~\ref{eq:se_rel} with the mass obtained from the analogous non-relativistic formula. 

\begin{figure}[!ht]
\vspace{0.75cm}
\includegraphics[width=8 cm]{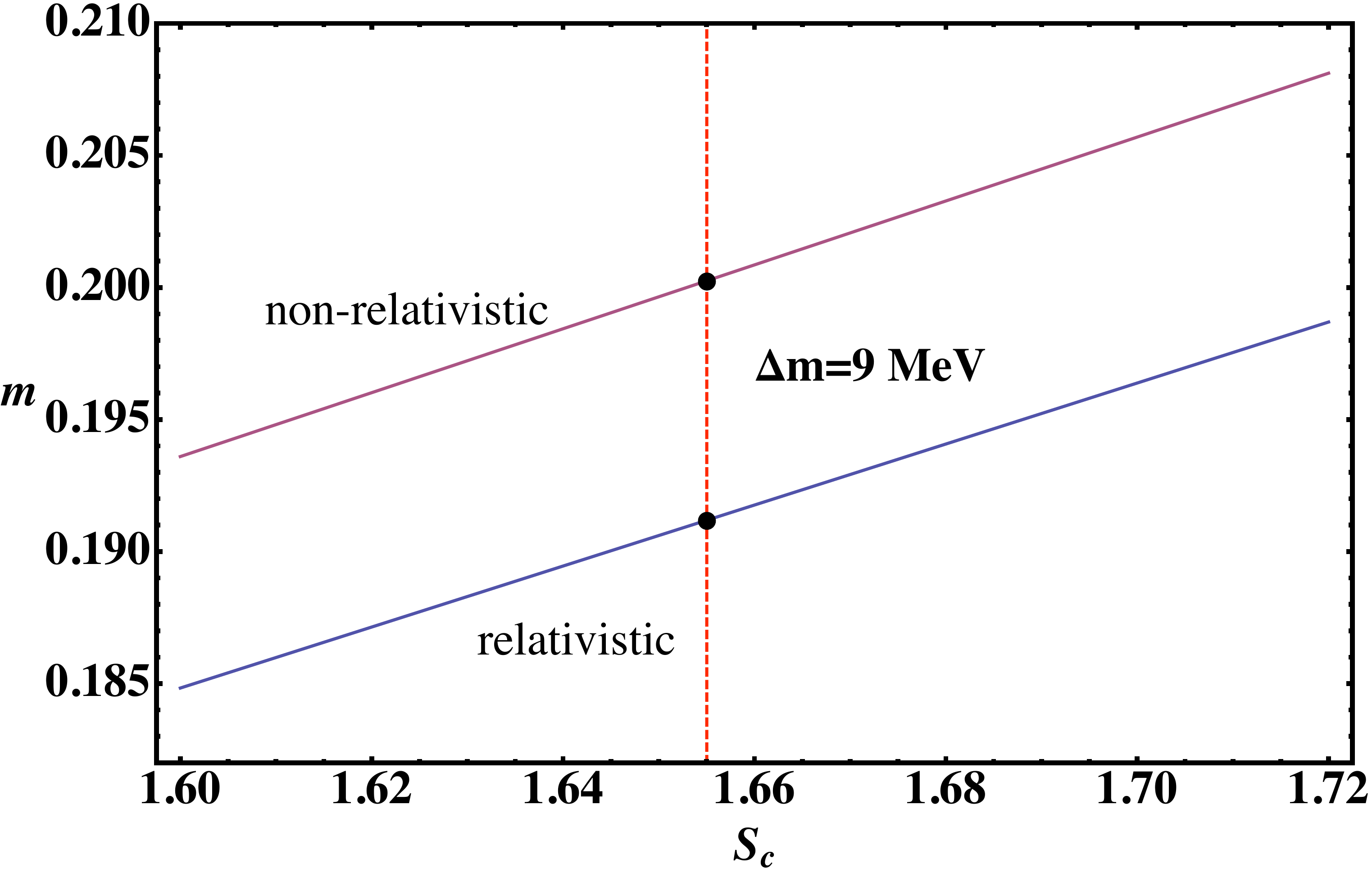}
\caption[h]{\label{fig:relcorr} Relativistic correction to the prediction of the monopole mass at $T=T_c$ without the inclusion of effects connected with inter-particle coulomb potential.}
\end{figure}

Fixing $T_c=0.27$ GeV and the jumping distance $\sqrt{2}d=\sqrt{2}n^{-1/3}$ we can extract the monopole mass as a function of the Feynman action. The result is presented in Fig.~\ref{fig:relcorr} where we can see that the relativistic correction is very small, in the order of $5\%$, therefore we can hope that, also introducing the potential, using the non-relativistic formula the error in the prediction of the mass remains small.

In order to include the potential we need an estimate of the coupling constant $g_m$. This estimate has been provided studying lattice correlation function in \cite{D'Alessandro:2007su,Liao:2008jg} to be such that plasma parameter
\be \Gamma={g_m^2\over 4\pi d T_c}\sim (1-2) \ee 

\begin{figure}[!ht]
\vspace{0.75cm}
\includegraphics[width=8 cm]{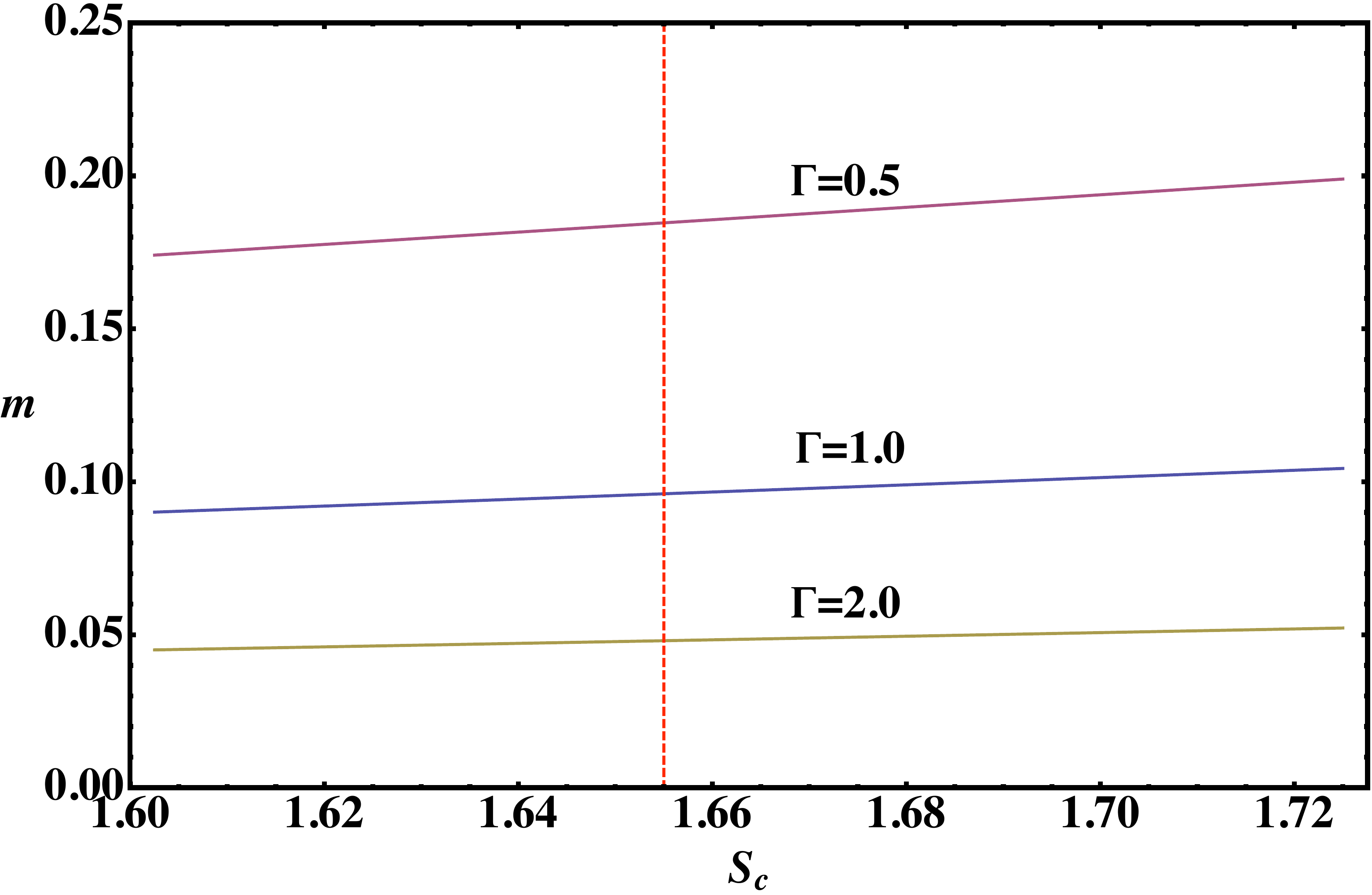}
\caption[h]{\label{fig:coulmass} Monopole mass prediction at $T=T_c$ using the non-relativistic formula for the Feynman parameter (Eq.~\ref{eq:se}). The three lines correspond to different strength of the coulomb potential.}
\end{figure}

In Fig.~\ref{fig:coulmass} you can see the dependence of the monopole mass from the Feynman parameter for three different choice of $\Gamma$: $\Gamma=0.5,1.0,2.0$. Remember that for $\Gamma\sim1$ the plasma is a liquid (only for $\Gamma<<1$ we have a gas and for $\Gamma\sim100$ a solid), therefore inside the range considered the state of the system does not change, nevertheless we can see that such a small variation in the plasma parameter is sufficient to produce a large variation in the monopole mass going from $50$~MeV for $\Gamma=2$ until $180$~MeV for $\Gamma=0.5$. On the other hand the dependence from the choice of the Feynman parameter is very small. 

In conclusion of this section we could say that in order to study the temperature dependence of the monopole mass it seems to be crucial to fix accurately the strength of the inter-particle potential.

\subsection{Discussion}
Liquid ${}^4$He at the lambda-point shows characteristic infinitely high peak in its specific heat, famously attributed by Feynman to divergence in the sum over the contribution of the k-polygons.  

The pure gauge theory deconfinement is second order for SU(2) group, it is the first order transition for $SU(N_c>2)$ and so far is designated to be a crossover one in QCD, with dynamical quarks. This means that added quarks  somehow tame the peak in the specific heat, to still large but finite hight. (It may still be a finite-volume effect.) It is perhaps fair to say that at the moment nobody has a clue why is it so, at least we are not aware of any definite ideas. 

The real-world QCD has fundamental quarks, and for those another transition -- chiral symmetry restoration -- coincides or is very close to deconfinement. However for different quarks, with adjoint (roughly twice larger) color charge, those are separated and deconfinement is the first order transition. 

In this paper we assumed that deconfinement is the BEC of monopoles, and it is crucially important to test on the lattice whether it is indeed true or not. This means in general to look for monopole condensate at $T<T_c$. We would suggest to follow Feynman and testing his predictions: (i) at $T_c$ the probability of polygons of any k gets comparable; and (ii) the extra action for ``jumps'' reaches the same critical value as for all other BEC's.  We know from simulations for ${}^4$He that those were indeed true in this case: why should monopoles be any different?

\section*{Acknowledgments}
We are greatly indebted to Massimo D'Elia, who had shared with us his published and unpublished results on lattice monopoles: this paper would never be finished without his data and encouragements. The work of MC is supported in part by the physics department of the University of Trento and by BMBF, GSI and the DFG Excellence Cluster ``Origin and Structure of the Universe''. The work of ES is supported in parts by the US-DOE grant DE-FG-88ER40388.

Note added: When this paper was completed, we learned about one more interesting application of strongly coupled charged
plasma possibly undergoing Bose-Einstein condensation:  helium dwarf stars, see
\cite{Gabadadze:2008mx}.

\end{document}